\documentclass{iopart}

\usepackage{epsfig}

\begin{document}

\title{Partial delocalization of two-component condensates in optical lattices}

\author{H A Cruz$^{1,2}$, V A Brazhnyi$^2$, V V Konotop$^{1,2}$}

\address{$^{1}$Centro de F\'{\i}sica Te\'orica e Computacional,
Universidade de Lisboa, Complexo Interdisciplinar, Avenida
Professor Gama Pinto 2, Lisboa 1649-003, Portugal}

\address{$^2$Departamento de F\'{\i}sica,
Universidade de Lisboa, Campo Grande, Ed. C8, Piso 6, Lisboa
1749-016, Portugal}

\begin{abstract}
We study management of localized modes in two-component (spinor) Bose-Einstein condensates embedded in optical lattices by changing interspecies interactions. By numerical integration of the coupled Gross-Pitaevskii equations, we find three different regimes of the delocalizing transition: i) the partial delocalization when the chemical potential of one of the components collapses with a gap edge and the respective component transforms into a Bloch state, while the other component remains localized; 
ii) the partial delocalization as consequence of instability of one of the components; and 
iii) the situation where the vector soliton reaches limits of the existence domain. 
It is shown that there exists a critical value for interspecies scattering length, below which solutions can be manipulated and above which one of the components is irreversibly destroyed.
\end{abstract}


%03.75.Lm Tunneling, Josephson effect, Bose-Einstein condensates in periodic potentials, solitons, vortices, and topological excitations 
%03.75.Mn Multicomponent condensates; spinor condensates 

%\submitto{\JPG}
%\maketitle

%\section{Introduction}

Management of matter waves in optical lattices (OLs) can be implemented in different ways, including variation of parameters of the periodic potential in time or change of the nonlinearity with help of the Feshbach resonance. 
In particular, by changing the amplitude of the lattice one can achieve the delocalizing transition, which consists in transformation of an initially localized mode, also referred to as a gap soliton, into a spreading out atomic wave packet. Such a transition was predicted for two- and three-dimensional single-component Bose-Einstein condensates (BECs) ~\cite{KRB,BS} and can be observed by adiabatically decreasing the amplitude of the periodic potential below some value, with subsequent restoration of the initial lattice profile. 
If the amplitude of the lattice reaches some critical value, the initially localized wave packet cannot be recovered at the end of the experiment and one observes a spreading out condensate. 
In a recent paper~\cite{LocDeloc}, it has been shown that the delocalizing transition can be observed also in one-dimensional (1D) systems, with help of interplay between linear and nonlinear lattices. 
This suggestion was based on the fact that the transition occurs in a system where small amplitude gap solitons cannot exist~\cite{BS}, i.e. where a minimal atomic number is necessary for the creation of a localized mode. 
From the physical point of view, the transition is related to modulational stability of the Bloch states bordering either the semi-infinite gap in the spectrum of Bogoliubov phonons or a gap induced by the periodic potential ~\cite{LocDeloc}.

The described results naturally rise questions: Whether other 1D systems allowing for delocalizing transition exist? and Whether other physical mechanisms resulting in delocalizing transition exist?  A simple positive answer to the first question readily follows from the fact that the quintic nonlinear Schr\"odinger equation with a periodic potential requires a critical number of particles for the creation of a gap soliton~\cite{quintic}, and thus can undergo the delocalizing transition. 
The answer to the second question is also positive, and moreover it offers another 1D system allowing for a new type of delocalizing transition. 
To explain this, we recall that new factors affecting the existence of gap solitons appear when one considers multicomponent systems in periodic potentials, and, in particular, binary mixtures or spinor condensates loaded in OLs: 
a necessary condition for existence of the modes is that the chemical potentials of both components must belong to the respective phonon gaps~\cite{mixed_symmetry}. This condition may be violated being subjected to change of the system parameters, thus leading to destruction of two-component gap solitons. 
Generally speaking, in such a situation one does not necessarily expect complete delocalization of the condensate, because one of the chemical potentials may still belong to a gap, while the other one reaches the gap edge. 
Instead, in this situation, one may observe a {\em partial delocalizing transition}, where one of the components spreads out, while the other one stays localized. 
Detailed description of this phenomenon constitutes the main goal of the present communication.   

Before going into details, we mention other aspects of the formulated problem. 
First, transformation between two different states of a condensate 
represents not only fundamental but also practical interest, as it allows for selective and thus controllable separation of the initially localized components of mixtures of BECs. 
Its mathematical relevance is determined by the fact that the transition allows for experimental study of the limits  of the existence of localized states.

Second, on the basis of preceding studies, the most direct approach to obtain the delocalizing transition would be the change of the parameters of the OL. 
However (this was checked by our extensive numerical simulations), the response of a binary mixture to such change is very similar to the one displayed by one-component BECs, and does not lead to the delocalizing transition. 
In particular, decreasing (increasing) of the potential results in expansion (narrowing) of vector gap solitons (c.f.  \cite{KRB,BS}). 
Essentially new phenomena can be found when management is performed by means of changing the interspecies interactions (for one-component BECs, the delocalizing transition induced by nonlinear management was suggested in~\cite{LocDeloc}). 

Third, it was also verified numerically that the delocalizing transition reveals qualitatively the same behavior for binary mixtures of two different species of atoms and for spinor condensates.
Therefore, in the present work we concentrate on the more simple dynamics of a spinor condensate induced by change of the interspecies interaction. 

Finally, since adiabatic change of the parameters in practice represents a particular type of management of mixtures of BECs, it is relevant to mention that, in Ref. \cite{2comp}, it was shown that interplay between the OL potential and the scattering length can assist building two-component gap solitons. 

%The paper is organized as follows. 
%In Sec. \ref{model}, we introduce the model and explain how solutions will be constructed. Sec. \ref{ccj_maior0} is devoted to the cases where all nonlinear coefficients have the same sign and where a delocalization transition exists. 
%In Sec. \ref{ccj_menor0}, nonlinear coefficients have different signs and regions of existence of gap solitons are observed. Sec. \ref{conclusions} our results are summarized.

\paragraph{The model and numerical method.}
%\label{model}

A diluted binary mixture of $F = 1/2$ spinor BECs in the mean-field approximation in a quasi-1D geometry is described by the coupled Gross-Pitaevskii (GP) equations
%(see e.g. \cite{mixed_symmetry} and references therein):     
\begin{equation}
i\frac{\partial\psi_{j}}{\partial t} = -\frac{\partial^2 \psi_{j}}{\partial x^2} - V\cos(2x)\psi_{j} %\nonumber \\
+ \left(\chi_{j}|\psi_{j}|^2 + \chi |\psi_{3-j}|^2 \right) \psi_{j}.
\label{GP}
\end{equation}
Here, $j=1,2$, and scaling of the variables is the same as in \cite{mixed_symmetry}. In particular, in the chosen units, the effective number of particles of component $j$ is calculated as $N_j=\int|\psi_{j}|^2dx$ and is related to the real number of atoms, ${\cal N}_j$, by the formula $N_j= g {\cal N}_j$, where 
$g\sim  10^{-3} \div 10^{-4}$ in the real experimental situation. The amplitude of the OL, $V$, is measured in the units of the recoil energy. The coefficients $\chi_j$ and $\chi$ characterize two-body interactions.

Since we are interested in a situation where $\chi=\chi(t)$ is adiabatically changing in time, stationary localized solutions of Eq.~(\ref{GP}) play a prominent role. We thus make an ansatz $\psi_{j}(x,t) = \Psi_{j}(x)e^{-i\mu_j t}$, where $\mu_{j}$ is the chemical potential of the $j^{th}$ component, and substituting in (\ref{GP}) we obtain ($j=1,2$)
\begin{equation}
\label{stationary}
\mu_{j}\Psi_{j} = - \frac{d^2 \Psi_{j}}{d x^2}  - V\cos(2x)\Psi_{j} +
%\nonumber \\
%&+&  
\left(\chi_{j}|\Psi_{j}|^2 + \chi|\Psi_{3-j}|^2 \right)\Psi_{j}.
\end{equation}

A diversity of localized modes supported by Eq.~(\ref{stationary}) possessing a given symmetry were described in \cite{mixed_symmetry}.  Here we focus only on the particular case of symmetric (even) modes. 
Moreover, to simplify the construction of the initial states (we do this by the shooting method), as well as for their systematic exploration, we restrict the consideration to the case where $(\chi_0 - \chi_1)(\chi_0 - \chi_2)>0$, where $\chi_0$ is defined as $\chi_0 = \chi(t=0)$. 
In this case, the simplest initial profile of a localized mode is characterized by equal chemical potentials $\mu_1 = \mu_2=\mu_0$. 
Then  $\Psi_j=\alpha_j^{-1}\Phi$, where $\alpha_j=\sqrt{|\chi_0^2-\chi_1\chi_2|/|\chi_0-\chi_{3-j}|}$ and the real function $\Phi$ solves the equation
\begin{eqnarray}
\label{phi}
\mu\Phi = - \Phi_{xx}  - V\cos(2x)\Phi + \sigma \Phi^3
\end{eqnarray} 
with $\sigma=$sgn$[(\chi_0^2-\chi_1\chi_2)(\chi_0-\chi_{3-j})]$ (due to the above constrain, it does not depend on $j$).
A relation among the numbers of particles is given by
%\begin{eqnarray}
%\label{N-relation}
$
	\int |\Phi|^2dx =N=\alpha_1^2N_1=\alpha_2^2N_2.
	$
%\end{eqnarray}

For each of the components, the OL results in a band spectrum, and chemical potentials must belong to the respective gaps. We further restrict our analysis to $\mu_0$ belonging either to the semi-infinite or to the first lowest gap. To be specific, we will also fix $V = 1$, which does not affect qualitatively the phenomenon.

As we investigate the behavior of localized modes subjected to the adiabatic variation of $\chi$, we first employ a Newton-Raphson (NR) method for constructing the branches of solutions, which ``depart" from the given $\Psi_j$ and are constrained to fixed values of $N_{1,2}$.  
When the strength of the interspecies interactions is changed, the chemical potentials of both components are changed as well, i.e. $\mu_j=\mu_j(\chi)$, provided that  $\mu_1(\chi_0)=\mu_2(\chi_0)=\mu_0$, and can be computed using the formula ($j=1,2$)
\begin{eqnarray}
\label{chemical}
\mu_{j} = \frac{1}{N_j} \int_{-\infty}^{\infty} \left(\left|\frac{\partial \psi_{j}}{\partial x}\right|^2  \right.&-& V\cos(2x) |\psi_{j}|^2 
\nonumber \\
&+&  \left. \chi_{j}|\psi_{j}|^4 + \chi|\psi_{3-j}|^2 |\psi_{j}|^2 \right) dx.
\end{eqnarray}

This formula is also used to calculate ``instant" chemical potentials in the temporal adiabatic dynamics, when $\chi=\chi(t)$, which is studied using the procedure as follows. 
For a given $\chi_0$, we start with the solutions $\Psi_j=\alpha_j^{-1}\Phi$ obtained from Eq.~(\ref{phi}). For the sake of definiteness, variation of the interspecies interaction is chosen as
\begin{eqnarray}
\chi(t) = (\chi_0 - \chi_m) \cos^2\left(\pi \frac{t}{T}\right) + \chi_m,
\label{ct}
\end{eqnarray}
with $\chi_m =\chi(T/2)$ and $T$ being the total time of simulations. In other words, we first increase (decrease) the nonlinearity until the maximal (minimal) value $\chi_m$ and then decrease (increase) it until the initial value $\chi_0$. Behavior of this system is investigated by a direct integration of Eq.~(\ref{GP}), with adiabatic variation of $\chi(t)$, as in Eq.~(\ref{ct}).

For management to be possible, the initial solutions must be experimentally feasible. Therefore we test the dynamical stability of the initial $\Psi_{1,2}$, by perturbing it with some addenda (in the presented results it is given by $0.05\cos(6.5 (x + \phi_j)) \Psi_j$, where $\phi_1 = 0.0$ and $\phi_2 = 0.3$) and numerically integrating Eq.~(\ref{GP}), with constant $\chi=\chi_0$.

\paragraph{Repulsive two-body interactions.}
%\label{ccj_maior0}

\begin{figure}[ht]
\centerline{\epsfig{file=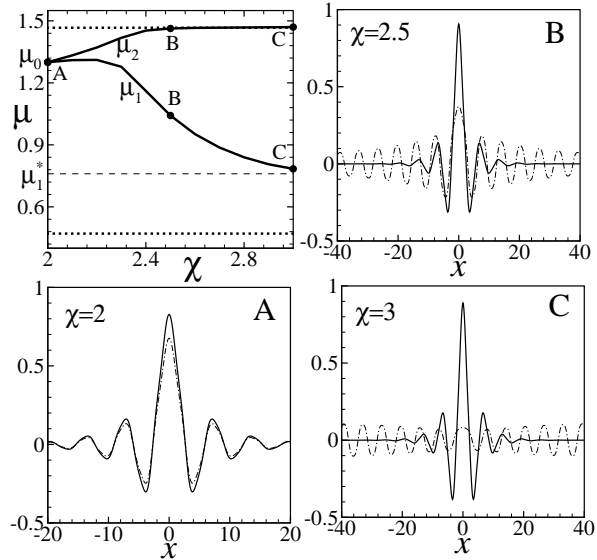, width=8cm}}
\caption{The chemical potentials {\it vs} nonlinear coefficient $\chi$ (left top panel), for $N_1 \approx 1.951$ and $N_2 \approx 1.303$. 
Dotted lines correspond to the limits of the first energy gap ($\mu \in [0.471; 1.467]$).
${\mu_1}^*$ is the chemical potential of the gap soliton of component $1$, with $N_1 \approx 1.951$, computed at $\Psi_2 = 0$ (see also Fig.~\ref{shoot_ccj_positivo}).
In the panels A, B and C, the profiles of stationary modes $\Psi_1$ (solid lines) and $\Psi_2$ (dashed lines) in the points A, B and C are shown. The initial chemical potential is $\mu_0 = 1.3$.}
\label{newt_ccj_positivo}
\end{figure}

First we consider a binary mixture with all repulsive interactions. 
To be specific we choose $\chi_1 = 1.0$, $\chi_2 = 0.5$ and $\chi_0 = 2.0$, what corresponds to $\alpha_1 = 1.53$, $\alpha_2 = 1.87$ and $\sigma=1$. 
We start with the existence of the stationary fundamental vector soliton for different $\chi$, using the NR method, as explained above. 
The chemical potentials $\mu_{1,2}(\chi)$ are depicted in Fig.~\ref{newt_ccj_positivo} (the left top panel), where one can observe two important features.

First, even infinitesimal change of $\chi=\chi_0+\delta \chi$ results in splitting of the chemical potentials: $\mu_j=\mu_0+\delta\mu_j$ with $\delta\mu_1\neq\delta\mu_2$. 
Second, the chemical potential of the second component monotonously grows with $\chi$ approaching the top of the first energy gap. The latter suggests that, at some $\chi = \chi_{cr}$ between points B and C (see the left top panel of Fig.~\ref{newt_ccj_positivo}), $\mu_2$ collapases with the gap edge.
For $\chi>\chi_{cr}$, the vector soliton with given numbers of particles in each component does not exist and thus, according to the arguments exposed above, $\chi_{cr}$ must determine the critical value above which the delocalizing transition occurs. 
Moreover, one can predict that this is a partial delocalization because the chemical potential of the first component still belongs to the gap and thus the respective component persists.
  
\begin{figure}[ht]
\centerline{\epsfig{file=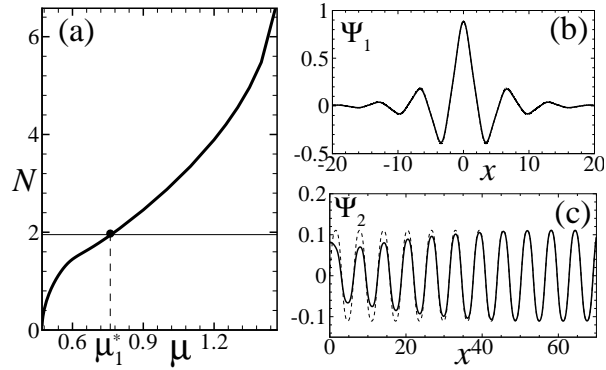, width=8cm}}
\caption{(a) The number of particles of the first component along the first energy gap ($\mu \in [0.471; 1.467]$), with $\Psi_2=0$. 
Horizontal thin line represents $N_1 \approx 1.951$ (i.e. the number of particles of the first component in the mixture analized in Fig.~\ref{newt_ccj_positivo}).
(b) The solid line represents the profile of $\Psi_1$ from Fig.~\ref{newt_ccj_positivo}C.  
The dashed line is obtained as a solution of Eqs.(\ref{stationary}), with $\Psi_2 = 0$ and $N_1 = 1.951$ (the lines are indistinguishable on the figure scale). 
(c) The solid line corresponds to $\Psi_2$ taken at the point C of Fig.~\ref{newt_ccj_positivo}, compared with the Bloch wave (the dashed line) calculated at the lower edge  of the second band ($\mu=1.467$) and having an amplitude matching the solution $\Psi_2$.}
\label{shoot_ccj_positivo}
\end{figure}

Although we use a rather large spatial domain of calculation $x\in [-48\pi$, $48\pi]$, the provided analysis of the stationary solutions is not yet conclusive about the existence of the modes, because of boundaries possibly affecting the expanding second component. 
Indeed, from the profiles of $\Psi_j$ shown in Fig.~\ref{newt_ccj_positivo}, one observes  tendency of the second component to the Bloch state of the upper band edge as $\chi$ exceeds  $\chi_{cr}$. 
To check numerically whether the delocalization is authentic, i.e. that the second component does not represent, say, a small amplitude gap soliton, affected by the boundaries, we recall that $|\Psi_2|^2$ becomes small enough when $\chi$ tends to $\chi_{cr}$. Hence the first component approximately solves Eq.~(\ref{stationary}) with $\Psi_2= 0$. That is why, in Fig.~\ref{shoot_ccj_positivo}, we constructed the lowest branch of the localized modes of the first component in the absence of the second component. 
For a given number of particles $N_1$, we found the chemical potential $\mu_1^*$ and constructed the one-component localized mode.

In Fig.~\ref{newt_ccj_positivo}(left top panel), one can see that, indeed, $\mu_1 \rightarrow {\mu_1}^*$ (see the dashed line) in the domain where $\mu_2$ collapses with the second band. 
Direct comparison of the explicit shapes of the localized modes in the single-component BEC and the first component in the binary mixture is shown in Fig.~\ref{shoot_ccj_positivo}(b), illustrating that the distribution of the first component is transformed into the one-component localized mode. In the same point, the second component becomes an extended Bloch state, as shown in Fig.~\ref{shoot_ccj_positivo}(c).

\begin{figure}[ht]
\centerline{\epsfig{file=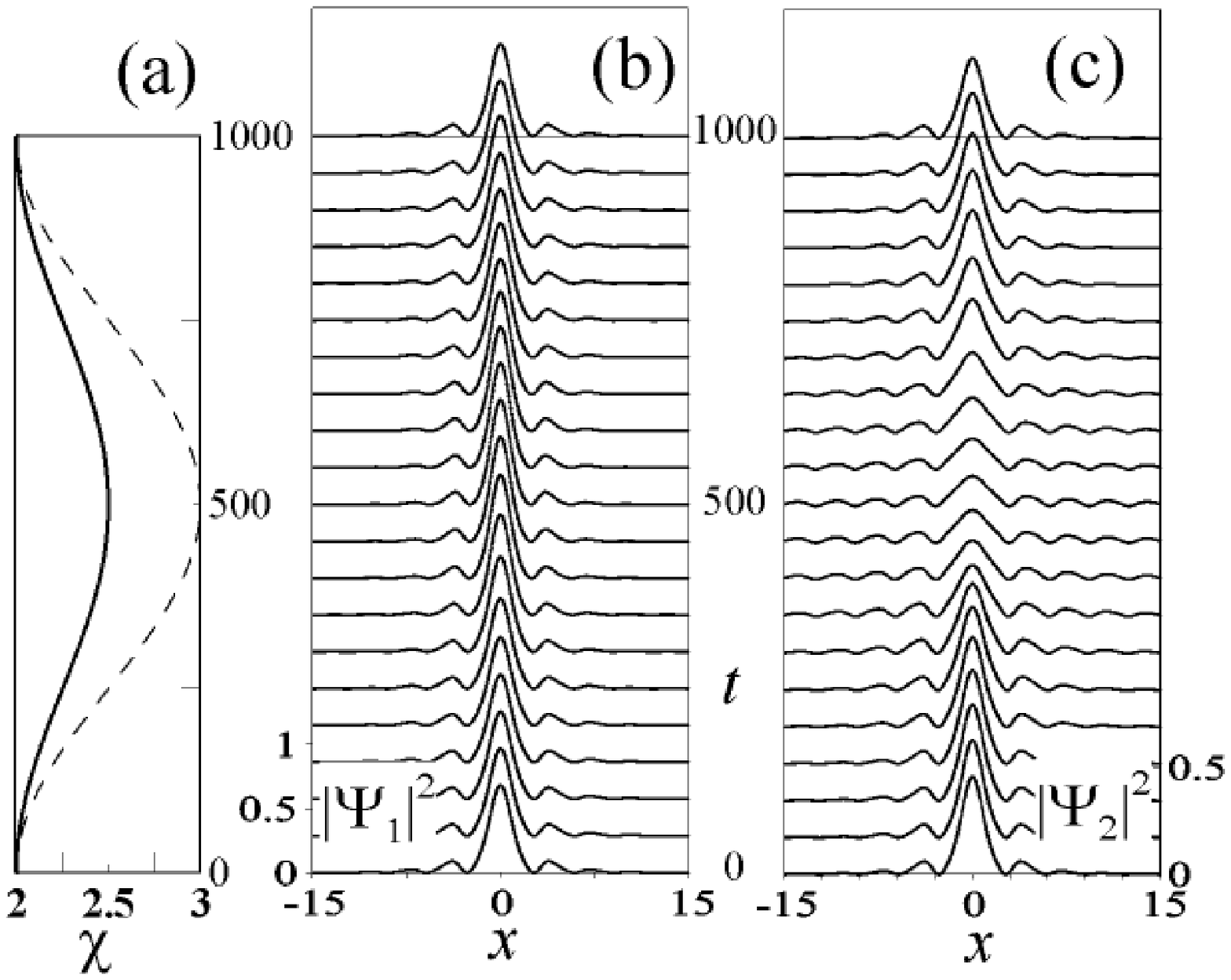, height=4cm}\epsfig{file=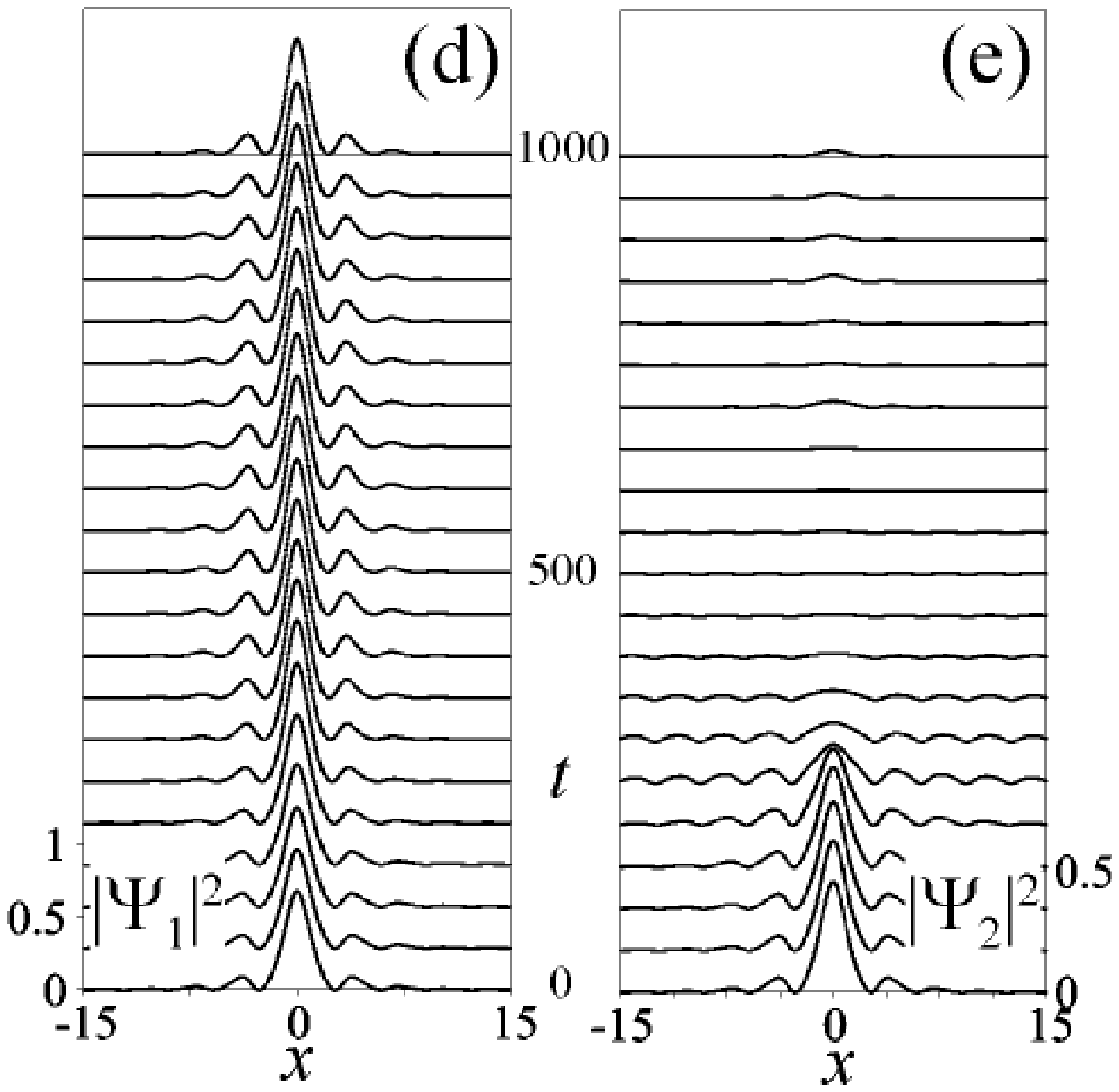, height=4cm}}
\caption{In (a), time dependence of the intraspecies interaction is shown. Solid and dashes lines correspond to $\chi_m=2.5$ and $\chi_m=3$.
Dynamics of the densities of the first component in (b), (d) and of the second component in (c), (e) are presented. $\chi_m = 2.5$ for [(b),(c)], and $\chi_m = 3.0$ for [(d),(e)]. Here, $\mu_0 = 1.3$ and $\chi_0 = 2.0$.}
\label{din_c25}
\end{figure}

Finally, we turn to the dynamical experiment. The results are shown in Fig.~\ref{din_c25}. Scanning values of $\chi_m$ between those corresponding to the points B and C in Fig.~\ref{newt_ccj_positivo}, one observes complete restoring of the vector soliton if $\chi$ does not reach its critical value, which is found to be $\chi_{cr}\approx 2.6$, while partial delocalization of the binary mode is observed when $\chi_m>\chi_{cr}$ (as predicted by the stationary analysis). 

Additionally, we explored the case with all attractive interactions ($\chi,\chi_j<0$). Symmetric solutions are now located in the semi-infinite gap and, as $\chi$ is adiabatically increased, the chemical potentials of both components grow, approaching the gap edge. However, we did not observe the delocalizing transition for negative values of $\chi$. It occurred only for $\chi>0$, i.e. when $\chi_j \cdot \chi$ becomes negative. Therefore, next we focus on this case.

\paragraph{Cases $\chi_j \cdot \chi < 0$.}
%\label{ccj_menor0}

\begin{figure}[h]
\centerline{\epsfig{file=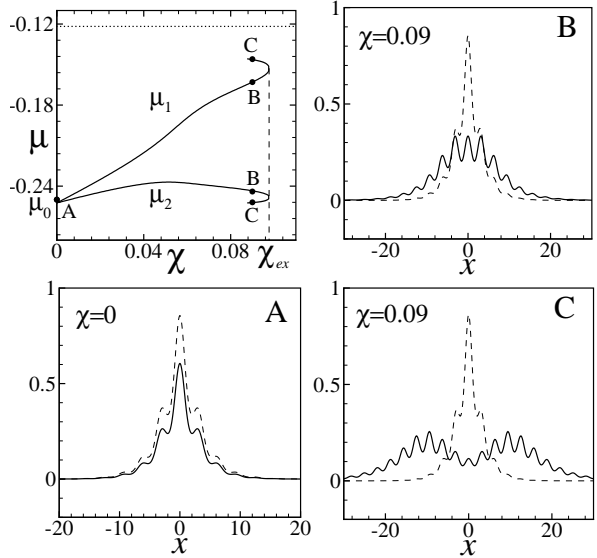, width=8cm}}
\caption{The chemical potentials {\it vs} nonlinear coefficient $\chi$ (left top panel). 
The dotted line corresponds to the upper edge of the semi-infinite gap. In A, B and C, the profiles of stationary modes $\Psi_1$ (solid lines) and $\Psi_2$ (dashed lines) are shown. Here, $N_1 \approx 0.996$ and $N_2 \approx 1.993$. The initial chemical potential is $\mu_0 = -0.25$. When $\chi$ approaches $\chi_{ex}$, solutions B and C bifurcate.}
\label{newt_w-025}
\end{figure}

Now we consider a binary mixture when intra- and interspecies interactions have different signs. 
To be specific, we investigate the case where $\chi_1 = -1.0$, $\chi_2 = -0.5$ and $\chi \geq 0$. In such a case, both chemical potentials belong to the semi-infinite gap.
Starting with uncoupled components, $\chi_0 = 0.0$, what corresponds to $\alpha_1 = 1.0$, $\alpha_2 = 0.71$ and $\sigma=-1$, we adiabatically increase $\chi$. 
Behavior of the chemical potentials $\mu_{1,2}(\chi)$ in the stationary problem and snapshots of the profiles of the components of vector solitons at different values of $\chi$ are presented in Fig.~\ref{newt_w-025}.

\begin{figure}[ht]
\epsfig{file=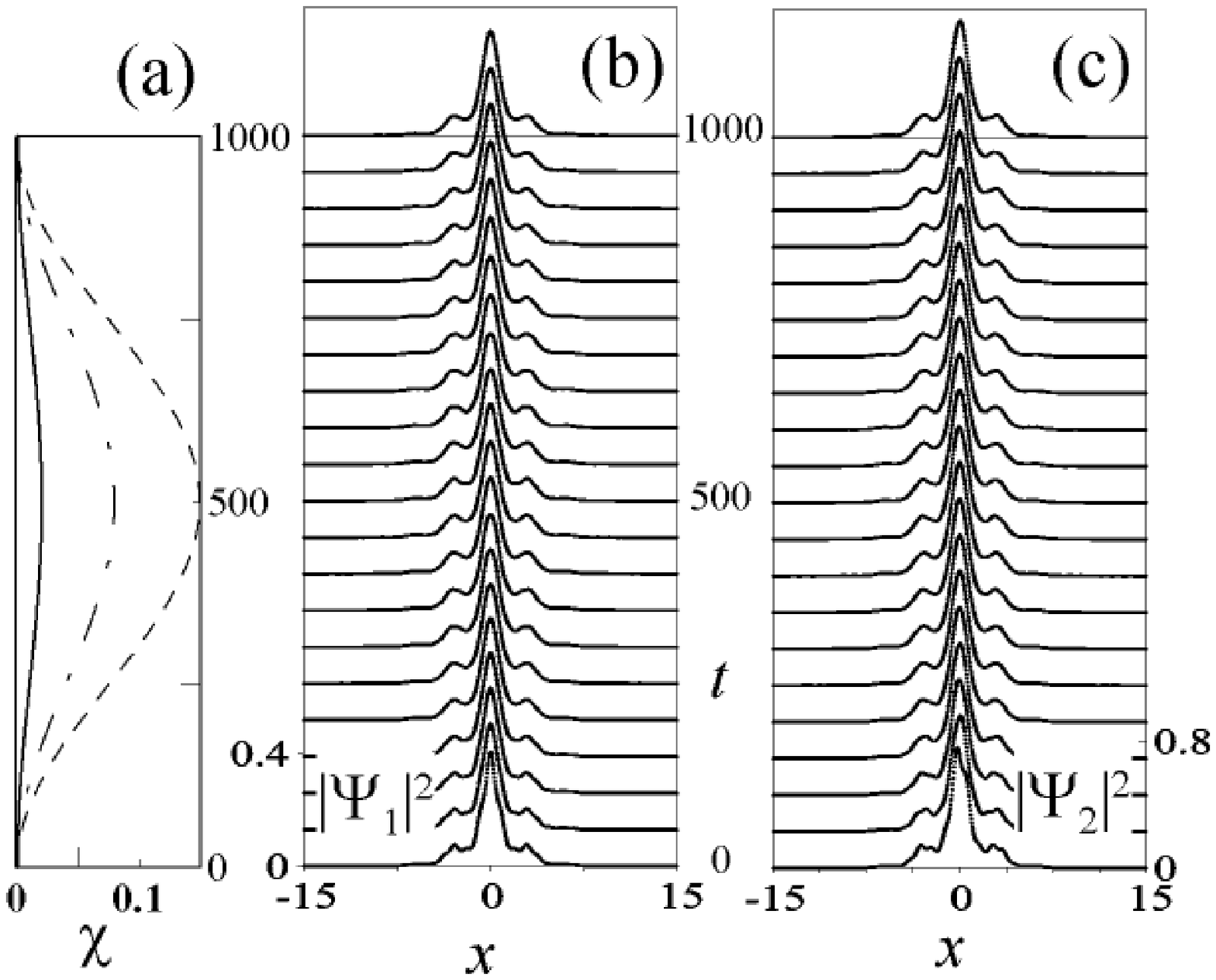, height=4cm}\epsfig{file=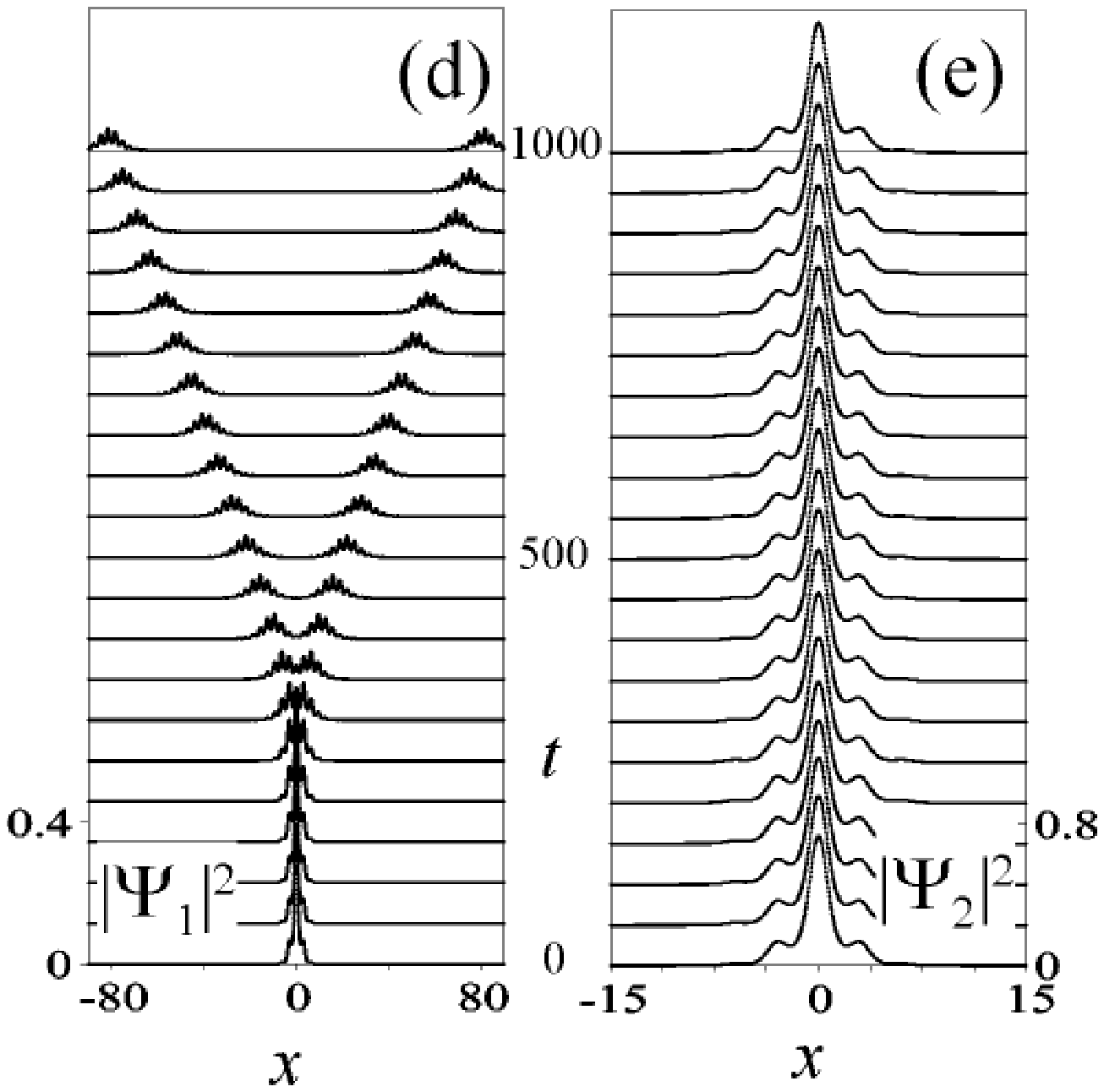, height=4cm}\epsfig{file=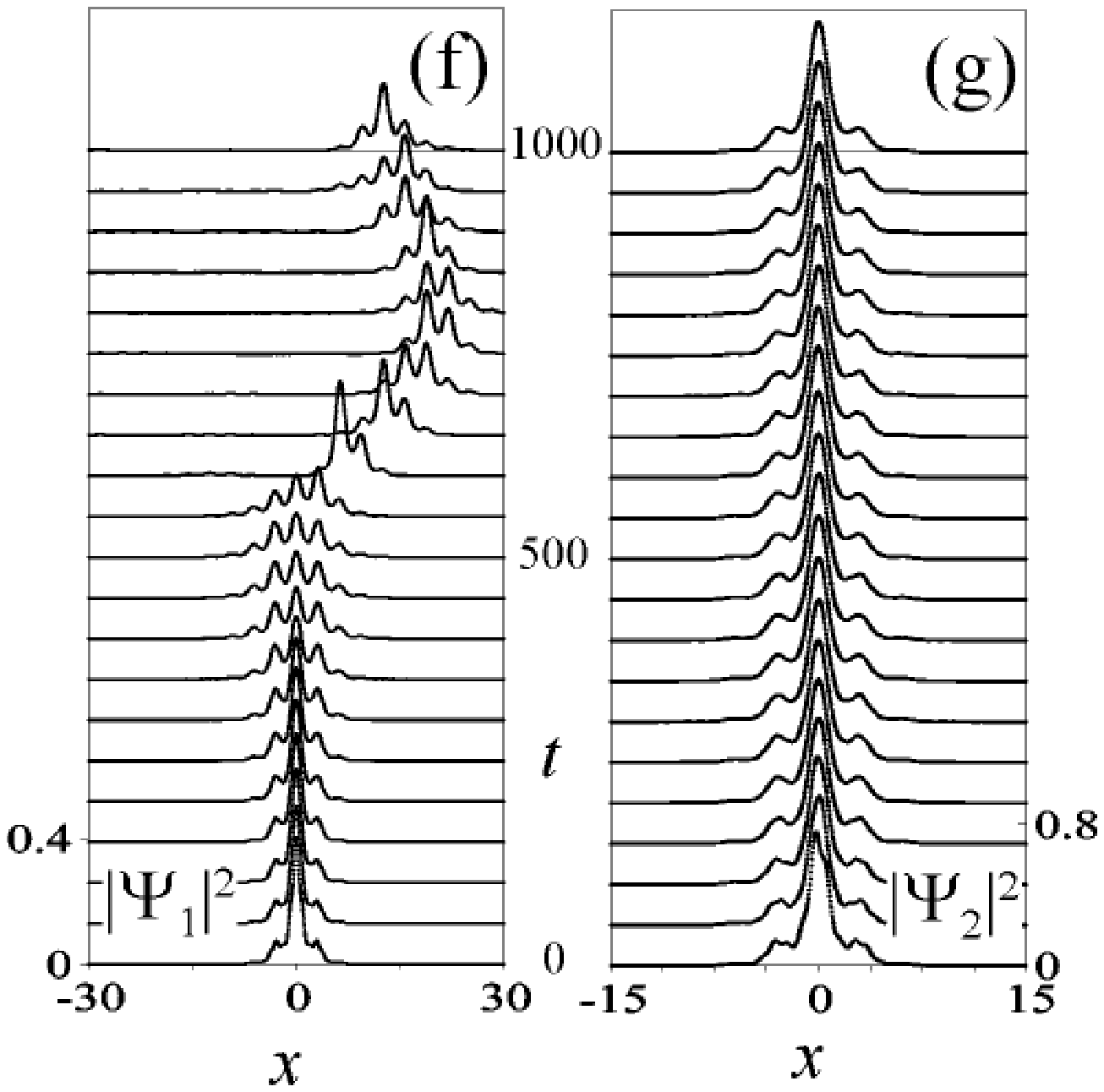, height=4cm}
\caption{In (a), the time dependence of the inter-species interaction is shown. Solid, dash-dotted and dashed lines correspond to $\chi_m=0.02$, $\chi_m=0.08$ and $\chi_m=0.15$.
Dynamics of the densities of the first component in (b), (d), (f) and of the second component in (c), (e), (g) are presented. 
$\chi_m = 0.02$ for [(b),(c)], $\chi_m = 0.15$ for [(d),(e)], and $\chi_m = 0.08$ for [(f),(g)]. Initial parameters are $\mu_0 = -0.25$ and $\chi_0 = 0.0$.
}
\label{din_w-025c008}
\end{figure}

As in the previous cases, change of $\chi$ results in splitting of chemical potentials $\mu_j$. 
Also, there exists a maximum strength for the repulsive interspecies interaction, $\chi=\chi_{ex}$, such that the fundamental vector soliton of the stationary problem can be found for $\chi<\chi_{ex}$. 
This point corresponds to the bifurcation point between two vector solitons whose profiles are represented in Fig.~\ref{newt_w-025}~B and C. Thus, contrary to the case of all repulsive interactions, now $\chi_{ex}$ is a bifurcation point where $\mu_{1,2}$ belong to the gap but where $\partial\mu_{1,2}/\partial\chi = \infty$. 
For the numbers of atoms as in Fig.~\ref{newt_w-025}, it was numerically found that $\chi_{ex} \approx 0.0975$. 
Passing to the dynamics, we integrated the GP equations (\ref{GP}). For different values of $\chi_m$ below the bifurcation point, $\chi_m < \chi_{ex}$, we found that the vector soliton can be restored to its initial shape, as shown in \ref{din_w-025c008}[(b),(c)]. 
However, if $\chi$  overcomes the critical value $\chi_{ex}$ (see Fig.~[\ref{din_w-025c008}(d), (e)]), one component (here $\Psi_1$) breaks into two repelling wave packets moving outwards the center, while the other component (here $\Psi_2$) transforms into a stable single-component localized mode. The observed behavior can also be interpreted as a kind of delocalizing transition. Its physical origin and, naturally, its manifestation are different from the ones reported in Fig.~\ref{newt_w-025}.

Additionally, instability of the vector soliton was observed for several values of $\chi_m<\chi_{ex}$. All such values are located in a region of the branch $\mu_2(\chi)$ between its inflection point and $\chi_{ex}$. We attribute this behavior to the existence of domains of instability along the branch $\mu_2(\chi)$. From Eq.~(\ref{ct}), we have that the velocity of $\chi(t)$ at the point $\chi_m$ is zero ($\dot\chi(\chi_m)=0$), therefore a mode spends long enough time in the vicinity of $\chi_m$ for the instability to develop. This is what happens in Fig.~\ref{din_w-025c008}[(f), (g)], where one of the components of the vector soliton (here $\Psi_1$) looses its stability, breaks from the second component and starts to move away from the center, while the second component maintains its shape. 
In the case where $\chi_m>\chi_{ex}$, presented in Fig.~\ref{din_w-025c008}[(d), (e)], although we pass in the instability region, we do not observe instability. This can be explained by that fact that the time spent by the vector soliton in the instability region was not enough for instability to start developing.

\paragraph{Conclusions}
%\label{conclusions}

We have studied the management of coupled gap solitons of a two-component spinor BEC embedded in an OL, by means of the adiabatic variation of the interspecies interaction. When all nonlinear interactions are repulsive, one of the components can be delocalized if subjected to a sufficiently large change of the interspecies scattering length, as its chemical potential collapses with the energy band edge, while the other component remains localized. On the other hand, if the intra- and inter-species interactions have different signs, partial delocalization is achieved when solutions reach limits of the existence domain. In this case, the manifestation of the phenomenon is different, as one of the components breaks in two gap solitons propagating outwards each other. Finally, we observed one more cause of the delocalizing transition: the instability of a vector soliton which develops more rapidly than the change of the scattering length occurs.

%\ack
H.A.C. acknowledges support of the FCT through the grant
SFRH/BD/23283/2005. V.A.B. was supported by the FCT grant
SFRH/BPD/5632/2001. 
The work was supported by the FCT and FEDER under the grant
POCI/FIS/56237/2004.


\section*{References}

\begin{thebibliography}{99}

\bibitem{KRB} Kalosakas G,  Rasmussen K \O and  Bishop A R 2002 {\it Phys. Rev. Lett.} {\bf 89} 030402

%\bibitem{AKKS} G.~L.~Alfimov, P. G. Kevrekidis, V.~V.~Konotop and M.Salerno, Phys. Rev. E {\bf 66}, %046608 (2002).

\bibitem{BS} Baizakov B B and Salerno M 2004 {\it Phys. Rev.} A {\bf 69} 013602

\bibitem{LocDeloc} Bludov Yu V, Brazhnyi V A and Konotop V V 2006 {\it Phys. Rev.} A {\bf 76} 023603

\bibitem{quintic} Sakaguchi H and  Malomed B A 2005 {\it Phys. Rev.} E {\bf 72} 046610;  Alfimov G L, Konotop V V and  Pacciani P 2007 {\it Phys. Rev.} A {\bf 75}  023624

\bibitem{mixed_symmetry}  Cruz H A,  Brazhnyi V A,  Konotop V V, Alfimov G L and Salerno M 2007 {\it Phys. Rev.} A {\bf 76} 013603

\bibitem{2comp}  Gubeskys A,  Malomed B A and  Merhasin I M 2006 {\it Phys. Rev.} A {\bf 73} 023607

%\bibitem{extcompI} L.D. Carr, and Y. Castin, Phys. Rev. A {\bf 66}, 063602 (2002).

%\bibitem{extcompII} F.K. Abdullaev, and M. Salerno, J. Phys. B {\bf 36}, 2851 (2003).

%\bibitem{delo_ring} J.-K. Xue, G.-Q. Li, and P. Peng, Phys. Lett. A {\bf 358}, 74-79 (2006).

%\bibitem{accelerate} V.A. Brazhnyi, V.V. Konotop, and V. Kuzmiak, Phys. Rev. A {\bf 70}, 043604 (2004).

%\bibitem{moving} P.G. Kevrekidis, D.J. Frantzeskakis, R. Carretero-Gonz\'{a}lez, B.A. Malomed, G. %Herring, and A.R. Bishop, Phys. Rev. A {\bf 71}, 023614 (2005).

%\bibitem{superlattice} M.A. Porter, P.G. Kevrekidis, R. Carretero-Gonz\'{a}lez, and D.J. Frantzeskakis, Phys. Lett. A {\bf 352}, 210 (2006).

%\bibitem{stabilizeI} B.B. Baizakov, V.V. Konotop, and M. Salerno, J. Phys. B {\bf 35}, 5105 (2002).

%\bibitem{stabilizeII} B.B. Baizakov, B.A. Malomed, and M. Salerno, Europhysics Lett. {\bf 63}, 642 %(2003).

 
%\bibitem{fesh_array} F.K. Abdullaev, E.N. Tsoy, B.A. Malomed, and R.A. Kraenkel, Phys. Rev. A {\bf 68}, %053606 (2003).

%\bibitem{fesh_managementI} F.K. Abdullaev, A.M. Kamchatnov, V.V. Konotop, and V.A. Brazhnyi, Phys. Rev. %Lett. {\bf 90}, 230402 (2003).

%\bibitem{fesh_managementII} V.A. Brazhnyi, and V.V. Konotop, Phys. Rev. A {\bf 72}, 033615 (2005).
%
%\bibitem{breather} P.G. Kevrekidis, G. Theocharis, D.J. Frantzeskakis, and B.A. Malomed, Phys. Rev. %Lett. {\bf 90}, 230401 (2003).

%\bibitem{fesh_molecular} R.S. Tasgal, G. Menabde, and Y.B. Band, Phys. Rev. A {\bf 74}, 053613 (2006).

%\bibitem{fesh_pre} M.A. Porter, M. Chugunova, and D.E. Pelinovsky, Phys. Rev. E {\bf 74}, 036610 (2006).
%
%\bibitem{quinticI} G.D. Montesinos, V.M. Perez-Garcia, P.J. Torres, Physica D {\bf 191}, 193 (2004).
%
%\bibitem{quinticII} F.Kh. Abdullaev, and J. Garnier, Phys. Rev. E {\bf 72}, 035603 (2005).
%
%\bibitem{pp} V.V. Konotop and P. Pacciani, Phys. Rev. Lett. {\bf 94}, 240405 (2005).

 


\end{thebibliography}
\end{document}